\documentclass[aps,onecolumn,superscriptaddress,11pt,nobibnotes]
{revtex4-2}
\DeclareUnicodeCharacter{02BC}{'}
\usepackage{amsmath,amssymb}
\usepackage{graphicx}% Include figure files
\usepackage{dcolumn}% Align table columns on decimal point
\usepackage{bm}% bold math
\usepackage[utf8]{inputenc}
\usepackage{braket}
\usepackage{float}
\usepackage{xcolor}
\usepackage{soul}
\usepackage{caption}
\usepackage{subcaption}

\usepackage{ragged2e}

\begin{document}
\preprint{APS/123-QED}
\title{Ultrafast Processing of Hyper-Entangled Bell States at the Optical Bandwidth Limit}% Force line breaks with \\
\author{Netanel P. Yaish} 
\affiliation{%
 Department of Physics and BINA Center for Nanotechnology,\\ Bar-Ilan University, Ramat Gan 5290002, Israel}
\author{Samata Gokhale}
\affiliation{%
 Department of Physics and BINA Center for Nanotechnology,\\ Bar-Ilan University, Ramat Gan 5290002, Israel}
\author{Avi Pe'er}
\email{avi.peer@biu.ac.il}
\affiliation{%
 Department of Physics and BINA Center for Nanotechnology,\\ Bar-Ilan University, Ramat Gan 5290002, Israel}
\date{\today}% It is always \today, today,

\begin{abstract}
\textbf {Abstract}: Bell states embody the most basic form of two-state entanglement and are a key component of quantum protocols of communication and sensing, indicating that high-speed processing of polarization Bell-states is highly desired for quantum technology. Yet, the processing speed of all current protocols is inherently limited by the electronic bandwidth of photo-detectors (Photomultiplier tubes, avalanche photo-diodes, etc.) that can process only $10^{6-7}$ photons/s, whereas standard broadband sources may easily produce $10^{10-14}$ photons/s. Using dual polarization, nonlinear SU(1,1) interferometry as an ultrafast quantum detector of the entangled bi-photons, we demonstrate the complete cycle of processing - generation, manipulation \textit{and detection} of polarization-entangled Bell states at an ultra-high photon-flux of $\sim\!5\times10^{11}$ photons/s ($\sim\!5$ orders of magnitude higher than standard methods), limited only by the optical bandwidth of the generating source. 
\end{abstract}

\maketitle

\begin{figure*}
    \centering
    \includegraphics[width=\textwidth, height=0.5\textheight, keepaspectratio]{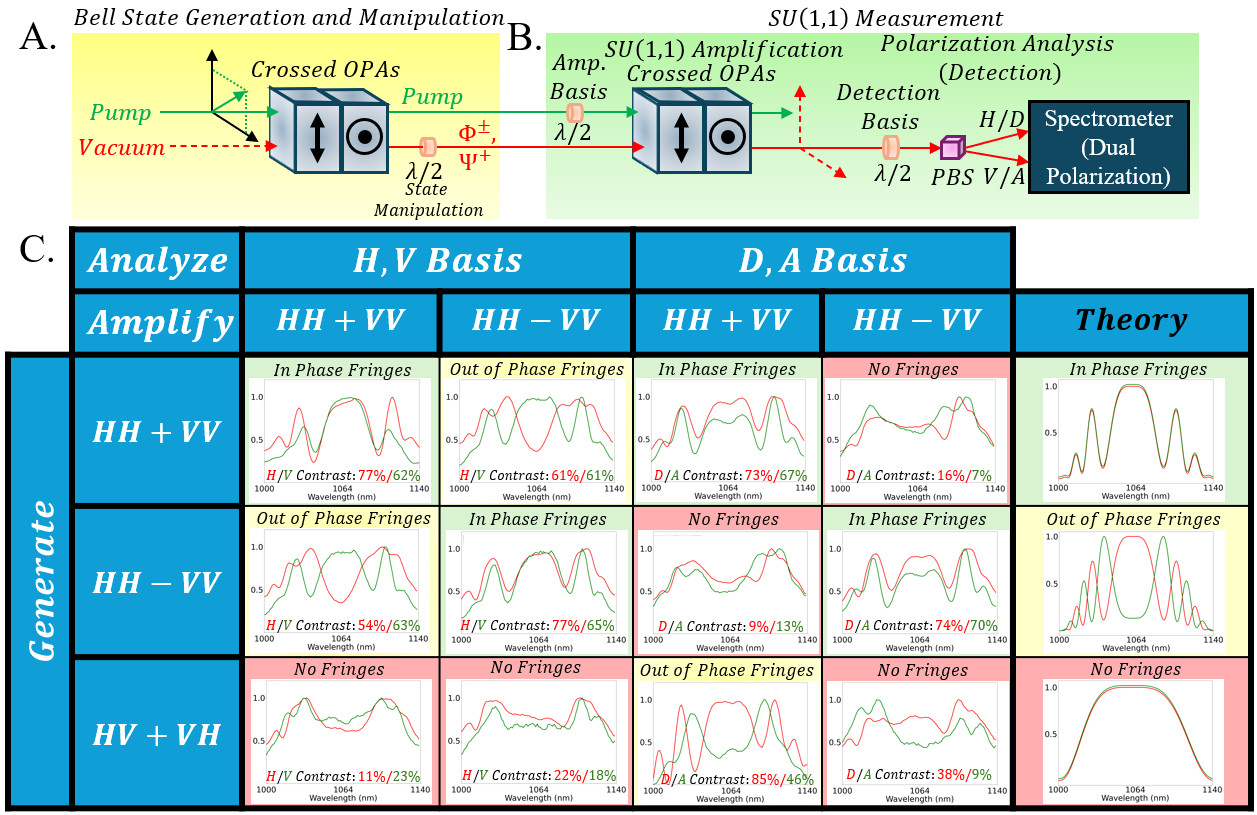}
    \captionsetup{font={stretch=0.95}}
    \caption{\justifying \textbf{Generation, manipulation and complete detection of Bell-states at an ultra-high flux - Concept and results.} \textbf{(A)} \textit{Generation and manipulation:} The $\Phi^\pm_{H,V}=|HH\rangle \pm |VV\rangle$ Bell states are generated by a pair of crossed type-0 nonlinear crystals, pumped by a common pump beam, polarized at $\pm45^0$. Each crystal acts as an Optical Parametric Amplifier (OPA) for one polarization, generating photon-pairs via SPDC over a broad bandwidth (limited by phase matching). The relative-phase between the polarization components of the Bell state $|HH\rangle \pm |VV\rangle$ is dictated by the polarization of the pump at $\pm45^0$. To manipulate the generated Bell-state, a $\lambda/2$ plate on the bi-photons beam can transform $\Phi^-_{H,V}$ into $\Psi^+_{H,V}=|HV\rangle +|VH\rangle$, but preserve $\Phi^+_{H,V}$, as explained in the text. \textbf{B}: \textit{The SU(1,1) measurement} is composed of an identical pair of crossed OPAs, followed by the simultaneous detection of the optical spectra of the bi-photons beams in both polarizations (either at the H/V or A/D basis). In the measurement OPAs, an incoming bi-photon can either be further amplified or annihilated, depending on the relative phase between the pair and the pump in each polarization. The amplified Bell state of the SU(1,1) measurement (one of $|HH\rangle \pm |VV\rangle$) is selected by a $\lambda/2$ plate on the pump beam (similar to the generation OPAs). After the SU(1,1) amplification, the detection basis in chosen by a $\lambda/2$ plate on the output bi-photons, and the polarization components are separated by a polarizing beam splitter (PBS) for simultaneous detection using a dual-polarization spectrometer. \textbf{C} \textit{Results}: The table presents the measured spectra (graphs) along with the expected result (text)  for each of the generated Bell-states (lines) and for each configuration of amplification and detection bases (Columns). The title text in each cell indicates the expected interference pattern (high-contrast / no fringes and in-phase / out-of-phase). The content of each cell shows the measured spectrum of the horizontal (red) and vertical (green) polarization, along with their observed fringe-contrasts. The last column presents the theoretically calculated spectra and fringe patters, assuming 20\% internal loss within our dual-SU1,1 (a reasonable estimation for our experiment).}
    \label{fig:CCSU11+Table}
\end{figure*}

Bell states - the maximally-entangled states of a pair of two-level systems (e.g. photon's polarization, atomic spins, etc.) are a cornerstone of quantum science and technology\cite{zou2021quantum, steane1998quantum, samuels2019importance, pirandola2018advances}. The Bell states can be written as  
\begin{equation}
\begin{split}
    \Phi^{\bm{\pm}}_{ H,V}& = \frac{1}{\sqrt{2}} \left( |HH\rangle \pm |VV\rangle \right)\\   
    \Psi^{\bm{\pm}}_{ H,V}& = \frac{1}{\sqrt{2}} \left( |HV\rangle  \pm |VH\rangle \right),
    \label{Bell States}
\end{split}
\end{equation}
where for light $H/V$ represents the horizontal/vertical polarization, $ |XY\rangle$ denotes the two-photon state of polarizations $X$ (signal photon) and $Y$ (idler) and the subscripts denote the polarization basis. Beyond the fundamental importance of Bell states as canonical demonstrators of entanglement in quantum mechanics\cite{nielsen2010quantum}, they are also key to major applications in quantum technology\cite{pan2012multiphoton}, particularly for quantum communications\cite{scarani2009security}. Consequently, protocols for high-speed processing, i.e. generation, manipulation \textit{and measurement} of Bell states are a major goal in experimental quantum optics. 

The major challenge for high-capacity processing is the measurement. While methods are available for the generation of $>\!10^{12}$ bi-photons per second \cite{shaked2014observing}, and for ultrafast manipulation of the Bell states \cite{eldan2026science} (as we review hereon), the standard methods for quantum measurement suffer from a hard limit: the speed of standard photo-detectors is inherently limited by their electronic bandwidth \cite{huang2015continuous, appel2007electronic, shaked2018lifting}, which can reliably detect at most $10^{6-7}$ bi-photons per second. If this major detection-bottleneck can be lifted the processing speed of polarization Bell states can be increased \textit{by several orders of magnitude}. 

Previously, methods for ultrafast measurement of time-energy entangled bi-photons at a high flux were introduced, using the quantum SU(1,1) interference \cite{shaked2014observing, shaked2018lifting, nechushtan2021optimal}, which replace the electronic coincidence detection with a physical pair-matching process that detects the coincidence directly at the optical level with no bandwidth limitation \cite{shaked2014observing, hudelist2014quantum, shaked2018lifting, nechushtan2021optimal} (see a brief review of SU(1,1) detection hereon). SU(1,1) interference has indeed been applied for quantum sensing and communication using a high-flux of time-energy entangled photons \cite{diamanti2016practical, eldan2026science, michael2019squeezing, nechushtan2025quantum}.

In this paper we introduce hyper-SU(1,1) interference as a detector of hyper-entangled Bell-states - i.e. entangled in both polarization and time-energy. Our dual-polarization broadband SU(1,1) interferometer (see figure \ref{fig:CCSU11+Table}A,B and explanation hereon) embodies a simple and robust protocol to fully process polarization Bell-states at ultrafast speeds. In what follows, we first review the generation of a high-flux of Bell states and their manipulation at high speed. This part adapts known methods of Bell-state optics for high-speed operation within an SU(1,1) configuration. We then describe the hyper-SU(1,1) Bell-measurement and analyze it in detail. 

The ability to generate a high flux of entangled photon-pairs goes hand-in-hand with time-energy entanglement within each pair. Specifically, the theoretical maximum-flux of entangled pairs of photons (that can be considered as single pairs) is simply $F_{max}\!=\!1/\tau_d\!=\!\Delta\omega$, where $\tau_d$ is the photon's correlation time, inversely related to the energy bandwidth $\Delta\omega$ of the bi-photons source. Thus, the key to a high-photon flux is a broad bandwidth of the photons. On the other hand, to maintain coherence between the generated bi-photons over a long time $\tau_p$ requires the pump field to be narrowband as $\delta\omega\!=\!1/\tau_p$. This inherently entails strong time–energy entanglement, where the energy of each photon is undefined over a wide range (the bi-photons bandwidth), but the energy-sum of the pair is very well defined (the pump bandwidth). Thus, to maximize the capability of a single experiment to generate a high number of coherent entangled pairs, $N\!=\!\tau_p/\tau_d\!=\!\Delta\omega/\delta\omega$, one should maximize the bandwidth-ratio between the photon-pairs and the pump.    

In the lab, a high-flux of broad bi-photons is naturally generated by spontaneous parametric down-conversion (SPDC) with broadband phase matching that is pumped by a narrow (CW) laser, particularly in type-0 or type-I crystals. Previous experiments have already produced extreme time–energy entanglement, where $\tau_d$ was as short as a single optical cycle \cite{nechushtan2025quantum, xue2021ultrabright} and the pair generation rates reached $10^{10}-10^{12}$ bi-photons/s, highlighting their potential for high-speed quantum processing \cite{eldan2026science, franson1989bell}. In the experiment here, the coherence time of the photons was $\tau_d\!\sim\!100fs$, allowing to generate up to $F_{max}\approx 10^{13}$ photons/s \cite{roeder2024measurement, dayan2005nonlinear}. Thus, combining the polarization entanglement with time-energy entanglement will allow us to generate a high-flux of polarization Bell states, thereby forming hyper-entangled photons, where neither the polarization nor the energy/time of each photon is known, and yet, they are precisely correlated to its twin in both polarization and energy/time  \cite{li2016complete, deng2017quantum, barreiro2005generation}.

To generate a high flux ($\sim\!5\!\times\!10^{11}$ pairs/s) of \textit{hyper-entangled} Bell-states, we employ the common configuration of two crossed type-0 nonlinear crystals (optical parametric amplifiers - OPAs) in series, whose major axes are crossed at 90 degrees \cite{iskhakov2009generation, chen2018polarization}  (see figure \ref{fig:CCSU11+Table}A). In this case, one crystal generates broadband photon-pairs at horizontal polarization, and the other - vertical. Taking the interaction Hamiltonian in each crystal as $H\!=\!g\left(a^{\dagger}_{s}a^{\dagger}_{i}+ a_{s}a_{i} \right)$, where $g\!=\!\chi l\alpha_p$ is the parametric gain ($\chi l$ is the effective nonlinear coefficient of the crystal and $\alpha_p$ is the classical pump amplitude), we can write the output state from a single crystal as $\ket{\psi_{out}}\!=\!e^{iH}\ket{0}\!\approx\!\left(1+ga^{\dagger}_{s}a^{\dagger}_{i}\right)\ket{0}$, where we assumed the parametric gain to be small $g\!\ll\!1$, i.e. spontaneous emission of single bi-photons. By tuning the pump polarization relative to the crystal axes, one can tune both the generation rate of bi-photons in each polarization (crystal) and the relative phase between them. Specifically, when the pump is polarized at $\pm45^\circ$, the crystals will generate equal amplitudes for both polarizations with equal/opposite phase, thereby naturally producing the $|HH\rangle \pm|VV\rangle$ Bell states\cite{kwiat1999ultrabright}. Specifically, the generated state after the pair of crystals is: 
\begin{equation}
\begin{split}
\ket{\psi_{out}}&=\ket{0}+\frac{g}{\sqrt{2}}(a^{\dagger^2}_H\pm a^{\dagger^2}_V)|0\rangle\\&=
\ket{0}+\frac{g}{\sqrt{2}} \left( |HH\rangle \pm|VV\rangle \right)\\&=\ket{0}+g\Phi^{\bm{\pm}}_{H,V},
\label{SPDC}
\end{split}
\end{equation}
where $a^{\dagger}_{H(V)}$ is the creation operator for the horizontal (vertical) polarizations. Alternatively, one can also use type-II down conversion \cite{kim2001interferometric}, but the bandwidth and conversion efficiency of type-II crystals is considerably lower than that of type-0, resulting in a much lower photon flux per pump power \cite{steinlechner2014efficient}.

After the generation, it is common to manipulate the Bell state of the bi-photons by rotating their polarization with a $\lambda/2$ plate (see figure \ref{fig:CCSU11+Table}A). Specifically, a simple rotation by $45^{\circ}$ of the polarization transforms the $ \ket{HH}\!-\!\ket{VV}$ state into $ |HV\rangle +|VH\rangle$, whereas  the $ |HH\rangle +|VV\rangle$ state remains invariant under this rotation. Specifically, when $ |HH\rangle \pm|VV\rangle$ (defined in equation \ref{SPDC}) is analyzed in the Diagonal /Anti-diagonal basis at ${\pm45^\circ}$, it's operator representation transforms according to
 \begin{equation}
 \begin{split}    
     \begin{aligned}
         a^{\dagger^2}_H+a^{\dagger^2}_V&=\frac{1}{2}\left[\left(a^\dagger_D+a^\dagger_A\right)^2+\left(a^\dagger_D-a^\dagger_A\right)^2\right] \\&= a^{\dagger^2}_D+a^{\dagger^2}_A,\\
         \textbf{OR}\\
         a^{\dagger^2}_H-a^{\dagger^2}_V&=\frac{1}{2}\left[\left(a^\dagger_D+a^\dagger_A\right)^2-\left(a^\dagger_D-a^\dagger_A\right)^2\right] \\&= a^{\dagger}_Da^{\dagger}_A+a^{\dagger}_Aa^{\dagger}_D
     \end{aligned}
     \label{Basis Change}     
\end{split}
\end{equation}
which is simply the creation operator for $|DD\rangle +|AA\rangle$ (or $|DA\rangle +|AD\rangle$). Indicating that a simple polarization rotation by ${45^\circ}$ can both differentiate between $ |HH\rangle \pm|VV\rangle$ and also generate $|HV\rangle +|VH\rangle$ (we note that the transformation is bi-directional, $\Phi^-_{H,V}=\Psi^+_{D,A}$ and $\Phi^-_{D,A}=\Psi^+_{H,V}$). To perform such state manipulations at high speed, one can use a fast polarization modulator (Pockel's cell), which can operate at GHz bandwidths. Higher manipulation throughputs can be obtained by wavelength-multiplexing of the manipulation - i.e. separating the spectrum the bi-photons into separate frequency channels (e.g. with a grating-based spectral shaper \cite{weiner2011ultrafast}), and manipulating every channel independently, adapting the concept of  \cite{eldan2026science} for parallel processing of quantum information. Of course, the manipulation is not limited to polarization rotation, and other manipulations, like phase / amplitude modulation, etc. can be realized similarly.  

It is therefore straightforward to generate and manipulate all the triplet states using this crossed-crystals configuration. Indeed, crossed-crystals similar to the outline of figure \ref{fig:CCSU11+Table}A were used previously to generate polarization-entangled pairs (both directly \cite{iskhakov2009generation} and within a polarization Sagnac interferometer \cite{chen2018polarization}). Generation of the singlet $\ket{HV}\!-\!\ket{VH}$ state is slightly more difficult, since polarization manipulations alone cannot transform states from the triplet manifold to the singlet. Thus, generation of the singlet requires selective control of the phase of one photon at only one polarization \cite{dayan2007spectral, iskhakov2009generation}.  And yet, the detection rate in all previous works remained limited by the bandwidth of the coincidence detectors at $<10^7$ photons/s, 5-6 orders of magnitude lower than the maximum flux $F_{max}\!=\!1/\tau_d$ of bi-photons. Here an optical measurement based on SU(1,1) interference can lead to a drastic improvement.  

In an SU(1,1) interferometer, two nonlinear media are set in series with a shared pump, such that the first nonlinear medium generates bi-photons, which are subsequently annihilated or amplified in the second medium, depending on the relative phase between the pump and the bi-photons at the second medium. This pair-annihilation in the second medium is an exact time-reversal of the generation process in the first medium, which can succeed only if the state of the pair arrived to the second medium intact. Thus, the pair annihilation acts as a physical coincidence mechanism for the bi-photons, as discussed in \cite{shaked2018lifting}. To generalize this concept for hyper-entangled Bell states, we utilize the dual-polarization SU(1,1) presented in figure \ref{fig:CCSU11+Table}A,B, where the generated bi-photons from the first pair of polarization-crossed OPAs (figure \ref{fig:CCSU11+Table}A) are measured by a second pair of polarization-crossed OPAs (figure \ref{fig:CCSU11+Table}B), followed by a polarization-analyzer of the light spectrum. Specifically, the pump polarization in the 2nd pair sets the amplification basis to be either $HH+VV$ or $HH-VV$ (controlled by the half wave plate on the pump beam before the 2nd pair). The analyzer then detects simultaneously the spectra of the bi-photon beams at both polarizations (the polarization-basis can be chosen as either H/V or D/A), as shown in figure \ref{fig:CCSU11+Table}B. In all we collect interference spectra at 4 cases - two amplification bases and two polarization bases.

To understand the operation of this dual SU(1,1) interferometer, let us first consider the simplest case of no manipulations of the polarization within the interferometer (pump or PDC). The output state after the second pair of crossed OPAs will then be:
\begin{equation}
    \ket{\Psi_{out}} = \ket{0} + \frac{g}{\sqrt{2}}(1+e^{i\theta_H})\ket{HH}\pm\frac{g}{\sqrt{2}}(1+e^{i\theta_V})\ket{VV},
\end{equation}
where the sign between the two polarization states is determined by the polarization of the pump ($\pm45^0$) and $\theta_{H(V)}$ is the phase acquired within the interferometer by the bi-photons in polarization $H(V)$. Since $\theta_{H(V)}(\omega)$ may depend on the optical frequency $\omega$ due to chromatic dispersion of optical elements (or due to purposeful modulation of the spectral phase), this phase leads to the observation of interference fringes across the spectrum of the bi-photons. An entangled Bell-state is characterized by a \textit{deterministic} relative phase between the two polarization components that is stable in time. Thus, the Bell states can be identified by monitoring this phase in real time and observing the relation between the two spectral-interferograms of the two measured polarizations. 

Let us now analyze in detail the properties of this SU(1,1) measurement and how it can detect each of the triplet Bell states, as summarized in table \ref{fig:CCSU11+Table}C, based on the manipulations on the polarization of the PDC or pump light within the interferometer, and on the measurement results in the H/V or A/D bases. Each of the generated polarization states - $HH+VV$, $HH-VV$, and $HV+VH$ can be probed under two possible amplification bases ($HH+VV$ or $HH-VV$) and the output can be analyzed in two detection bases ($H/V$ or $D/A$) - as represented in the four columns of the table. 

We begin by considering the simple cases where the incoming Bell states are amplified and analyzed in the same $H/V$ basis. For $HH\pm VV$ incoming states, the entire system is effectively composed of two uncoupled SU(1,1) interferometers - one for each polarization, where the $H(V)$-detection shows the spectral fringes of the $H(V)$ interference. Thus, one expects high-contrast fringes to appear across the spectrum of both H and V. When the generation and amplification states are matched ($HH+VV/HH+VV$ or $HH-VV/HH-VV$), the observed fringes in $H$ and $V$ will be \textit{in phase} since the internal phase is common to both polarizations. Similarly, when the generation and amplification bases are switched ($HH+VV/HH-VV$ or $HH-VV/HH+VV$) we expect high-contrast fringes that are \textit{$180^\circ$ out-of-phase}, reflecting the $\pi$ phase shift of one polarization component due to the polarization rotation of the pump. In the table, the title of each cell indicates the expectation for the represented case, and the graphs within the cell show the simultaneously measured spectra of both $H$ (red) and $V$ (green) along with their interference contrasts. Indeed, the analysis of the $HH\pm VV$ states shows high contrast fringes that are either in phase or out of phase, as expected. 

In contrast, when the $HV+VH$ state enters the SU(1,1) measurement no spectral fringes are expected for H/V detection, since the input state does not match the  annihilated state of the SU(1,1) amplifiers. Intuitively this can be understood since now the signal and idler photons are necessarily split between the $H/V$ polarizations, indicating that each OPA will interact with only one photon of the pair, whereas SU(1,1) interference in type-0 crystals requires both photons to have the same polarization. Indeed, the experimental observation shows very low-contrast fringes for $HV+VH$ for both amplification states (the residual interference is due to experimental imperfections that we discuss later on) 

When the detection basis is rotated to $D/A$ the amplification basis transforms according to equation \ref{Basis Change}:  $HH-VV$ transforms into $DA+AD$, but $HH+VV$ is preserved as $DD+AA$. Thus, when the $HH\pm VV$ states enter the SU(1,1) measurement, high-contrast, \textit{in phase} fringes are expected \textit{only} when the generation and amplification states are matched (similar to the H/V analysis), but when the the generation and amplification states are switched, no fringes are expected to appear in any polarization, since the annihilated state cannot match the input state (one is $HV+VH$, while the other $HH+VV$). Indeed, the experiment observes high-contrast fringes for the matched cases, but only low-contrast residual interference for the switched cases.

A surprising result arises however when the generated $HV+VH$ is detected in the $D/A$ basis. Since the rotation to the A/D basis transforms $HV+VH$ back to $DD-AA$, we find that when the incoming $HV+VH$ enters into $HH+VV$ amplification, high-contrast \textit{out-of-phase} fringes are expected for $D/A$ detection, as indeed observed experimentally. This "revival" of the interference fringes is surprising since none of the SU(1,1) amplifiers (that still operate in H/V basis) show any interference (since each amplifier "sees" only one photon of the pair). Thus, the observed revival of high-contrast interference when we mix the $H/V$ polarizations in the $D/A$ analysis can only be the result of the polarization entanglement in both the generated and the amplified states.

\begin{figure}[hbtp]
\centering
    % Left Subfigure
    \begin{subfigure}[b]{0.55\textwidth}
        \centering
        \includegraphics[width=\columnwidth]{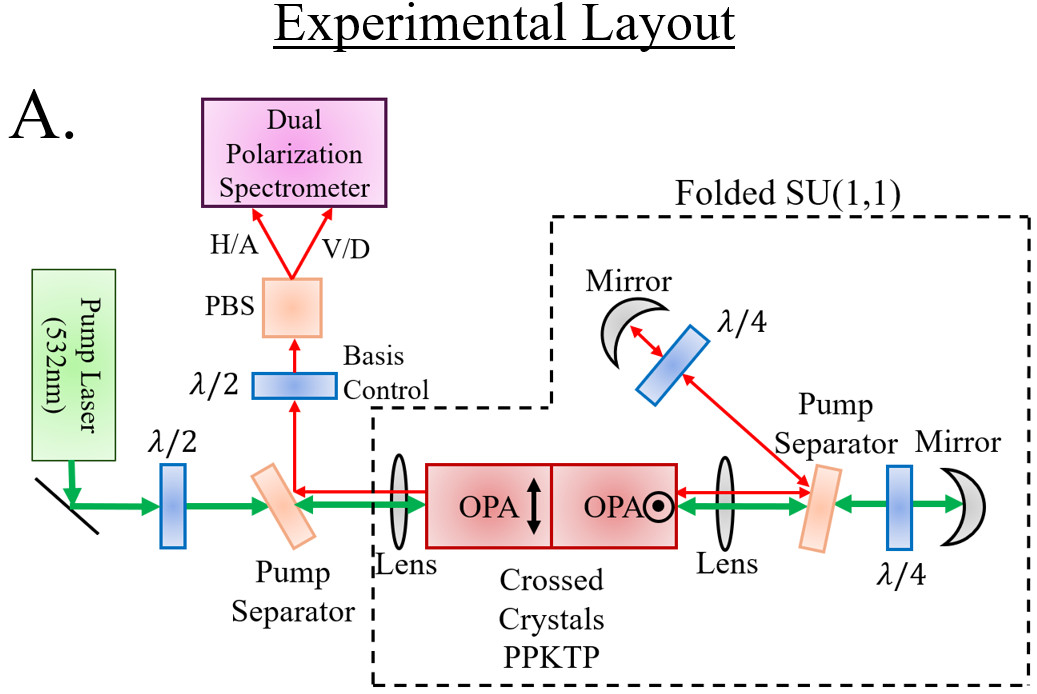}
        \phantomcaption
        \label{System}
    \end{subfigure}

    % Right Subfigure
    \begin{subfigure}[b]{0.55\textwidth}
        \centering
        \includegraphics[width=\columnwidth,height=0.9\textheight,keepaspectratio]{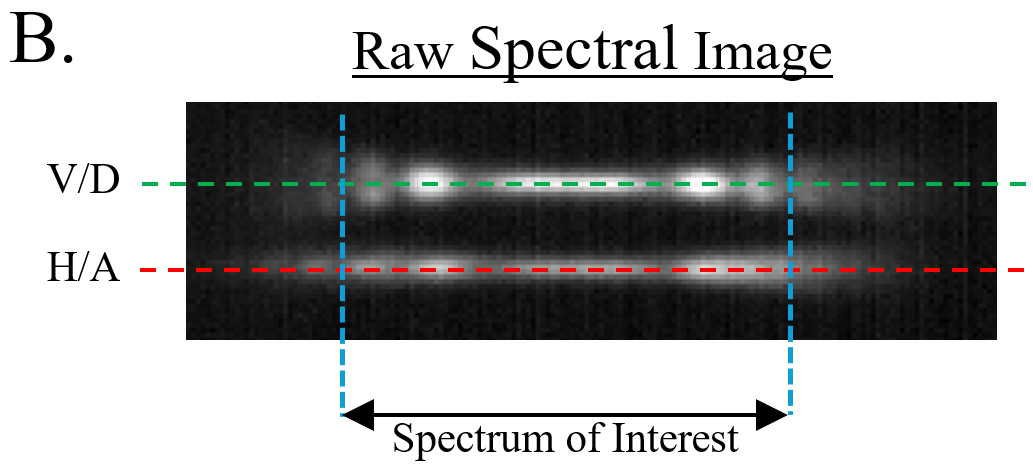}
        \phantomcaption
        \label{RD}
    \end{subfigure}
     \captionsetup{font={stretch=0.95}}
    \caption{\justifying \textbf{Experimental layout and measurement procedure. (A)} Experimental layout of the dual-polarization SU(1,1) in a folded configuration. The crossed-crystal OPAs (two 12mm-long, periodically-polled KTP crystals) are pumped by a CW, single-frequency laser at 532nm, linearly polarized at ${\pm45^\circ}$ relative to the crystal axis ($\sim750mW$, $\sim375mW$ per crystal), and both the pump and the generated bi-photons are retro-reflected into the same crystals for measurement (as shown). The pump beam was focused to the mid-point between the two crystals with a Rayleigh range that stretches across both crystals. In order to independently control the polarization Bell-state to be measured and the amplification Bell-state of the SU(1,1) measurement, we separate the paths of the pump and the down-converted bi-photons using a dichroic filter and independently control the polarizations of the pump and the bi-photons with separate $\lambda/2$ plates between the passes ($\lambda/4$ double-passed). After the measurement-OPAs, we use another $\lambda/2$ plate and a polarizing beam splitter (PBS) to select the detection basis ($H/V$ or $D/A$) and separate the polarizations for independent detection. Finally, we detect the intensity spectra of both polarizations simultaneously with a home-built dual-polarization spectrometer (composed of a wave plate, two Brewster prisms of SF11 glass in series and an InGaAs CCD camera, not shown). \textbf{(B)} Example of the raw CCD images, showing the captured spectra of both polarizations with correlated SU(1,1) fringes (in phase). During the detection we recorded a continuous video sample of the free-running spectral interference pattern (at 50 frames/s) and analyzed it to retrieve the interference contrast and phase -correlation. The integration time of each video frame was sufficiently short compared to the typical time of phase fluctuations in the free-running SU(1,1) interferometer, indicating that every frame represents an "instantaneous" spectral phase of both polarizations \textit{simultaneously}, allowing us to deduce the relative phase between them. }
\end{figure}

Figure \ref{System} shows the experimental design of the dual SU(1,1) interferometer, constructed in a folded configuration, where two crossed OPA crystals generate the hyper-entangled bi-photons across a $\sim\!70$nm wide spectrum around 1064nm ($\Delta\omega\!\approx\!10THz$ bandwidth) in the forward direction and then measure them in the backwards direction. Figure \ref{RD} shows an example of a raw captured CCD image that demonstrates the SU(1,1) interference fringes of both polarizations, allowing us to calculate the interference contrast of each polarization and the phase-correlation between them (deducing the relative-phase between the polarizations). The processed results are summarized in table \ref{fig:CCSU11+Table}C for the different generated Bell states ((rows) and the various amplification states and detection bases (columns). The parametric gain of each crystal is estimated to be $g\approx0.13\rightarrow g^2\approx0.019$, based on the estimation of the “equalization” pump power  $P_{eq}\approx15Watt$, where the stimulated generation of pairs equals the spontaneous generation, i.e. $G_{eq}=e^{g_{eq}}=2$ \cite{dayan2005nonlinear}.  Thus, at the pump level of the  experiment ($P\approx375mW$ per crystal) the gain is $g=g_{eq} \sqrt{P/P_{eq}}\approx0.13$.  Since the average photon number per Hz per second  is $N_s=sinh^2g\approx g^2$, the photon flux is estimated to be $F\!=\!g^2\Delta\omega \!\approx \!5\times10^{11}$ photons/s (for the total bandwidth of $10THz$).

To support the experiments with theory, we performed calculations of the expected measurement spectra for the detected polarizations, under realistic experimental parameters of parametric gain and internal loss. The calculations relied the standard SU(1,1) model \cite{ou2020quantum, gokhale2026quantum} while taking into account the polarization effects. Figure \ref{fig:CCSU11+Table} shows the calculated spectral interference patterns for three representative cases - in-phase interference, out-of-phase and no-fringes, indicating an expected fringe contrast of  $\sim90\%$ for the high-contrast cases, limited primarily by the internal losses, estimated at 10\%, whereas no contrast is predicted for states measured in orthogonal bases. Experimentally, all the expected high-contrast cases demonstrated $>50\%$ contrasts, whereas the no-fringe cases demonstrated low contrast between $>50\%$. The residual interference for the no-fringe cases, as well as the reduced visibility for the high-fringe cases is primarily due to imperfect polarization manipulations by the wave=plates (axis-errors and chromatic variation of the zero-order wave plates that were used), as well as residual non-collinearity and spatial mode-mismatch between the bi-photons and the pump beam during the 2nd pass through the crossed OPAs (SU(1,1) amplification). 

Our use of an SU(1,1) interferometer as our measurement device, means the efficiency of photon pairs produced is equivalent to the efficiency of generation. Therefore, the output of this measurement was then detected using an integrated CCD signal. At any given moment (time delta) all parallel pairs are detected simultaneously by our interferometer, such that the number of pairs is related to the spectrum generated. Due to the time-energy entangled aspect of our bi-photons, multiple pairs of the same frequency (energy) will not arrive simultaneously (i.e. within the given time delta), avoiding erroneous values in our measurement.

To extract the purity of the bi-photon Bell-states that we generate, we note that the visibility (interference contrast, $V=100\times(I_{max}-I_{min})/(I_{max}+I_{min})$) directly reflects the purity of the state by the relation $\eta(t)=|t|^2=V^2$ where $\eta$ is the purity of the quantum state - the fraction of entangled photons at a specific frequency out of the total number of photons at that specific frequency ($\eta(t)\equiv\langle N^{pairs} \rangle/\langle N^{photons} \rangle$) \cite{shaked2014observing, nechushtan2021optimal}. Therefore, to evaluate the visibility of each polarization (\ref{fig:CCSU11+Table}C) we post-selected the frames with the maximum (constructive, $\phi=0$) and minimum (destructive, $\phi=\pi$) intensity at the $1064nm$ point of the spectrum. Specifically, we selected the frames where the spectral fringes were at a maximum at $1064nm$ and compared to the spectral fringes of the frames with the minimum intensity at the $1064nm$ point. The entanglement between the polarization states is directly indicated by the fact that despite the inherent instability of the phase of each polarization (that was randomly drifting and fluctuating), the relative phase between the two polarizations remained stable at all times. 

In our  experiment, we observe the fringe pattern with a CCD camera at relatively long integration times (5-15ms), indicating that our measurement speed is the frame rate of the camera, which is very low relative to the photon-rate. Thus, the measured intensity of a given pixel reflects the \textit{average} number of photons per that pixel ($\gg\!1$), which simply indicates the probability to detect a photon at a given wavelength. Although the detection rate of a pixel is limited by its integration time, the total photon rate remains unlimited, since the measured spectrum is distributed over many pixels that detect  in parallel different spectral "channels". Thus, the total number of pixels is the lever that multiplies the single pixel rate, and one can always increase the total processing rate by spreading the spectrum over more pixels. With commercially available line-array cameras (2048 pixels with $1\mu s$ integration time) the processing rate can already reach  $>\!2$ GHz. Faster line-arrays of detectors can obviously push that rate much higher well into the THz regime. 

Note that the single-photon regime of the experiment is guaranteed by the low parametric gain of the OPAs  and is not affected at all by the integration time of the camera. Even when each pixel of the camera detects many photons during its integration time, the "coincidence" detection is performed physically by the SU(1,1) interferometer, which is inherently ultrafast (limited by the coherence time $\tau=1/\Delta\omega$). Thus, as long as the parametric gain is sufficiently low, the single-photon nature is assured. While the current experiment and discussion is aimed specifically at the low-power regime of single bi-photons, the SU(1,1) platform is directly applicable also well beyond single photons (e.g. for quantum sensing below the shot-noise limit \cite{gokhale2026quantum}. 

In conclusion, we demonstrated both the generation and complete quantum measurement of an ultra-high flux of hyper-entangled bi-photons in both polarization and time-energy. We also determined a distinct pattern uniquely identifying each of the triplet states. The ultrafast measurement of the entire flux is made possible by harnessing the nonlinear SU(1,1) interference as a physical two-photon detector. In the future, our concept can be used for applications in quantum metrology \cite{PhysRevA.97.010301} as well as quantum computing \cite{deng2017quantum}. For metrology, at a high parametric gain, the dual SU(1,1) interferometer can harness the generated hyper-entangled squeezed light to form a quantum sensor of birefringence with sub-shot-noise sensitivity \cite{gokhale2026quantum}. The Bell states generated can also be useful for quantum communication, where the large bandwidth can immediately translate to speed and the polarization entanglement can be used to help deterministically separate photons pairs in QKD experiments. Specifically, one can frequency-multiplex many communication channels simultaneously across the bi-photons spectrum \cite{eldan2026science}.  

\bibliographystyle{apsrev4-2}
\bibliography{NetanelPaper}% Produces the bibliography via BibTeX.
\end{document}